\documentclass[aps,pre,
   11pt,
   reprint,
   superscriptaddress,
   eqsecnum,
   showpacs,
   preprintnumbers,
   longbibliography,
   nofootinbib
   ]{revtex4-1}
\usepackage{color}
\usepackage{framed}
\usepackage{subfigure}
\usepackage{time}
\usepackage{graphicx}
\DeclareSymbolFont{AMSb}{U}{msb}{m}{n}
\DeclareSymbolFontAlphabet{\mathbb}{AMSb}
\usepackage{physics}                           
\newcommand{\ef}[1]{\eqref{#1}}                

\newcommand{\mc}[1]{\mathcal{#1}}              
\newcommand{\etal}{\textit{et~al.}}        
\newcommand{\vs}{vs.}                      
\newcommand{\Schrodinger}{Schr{\"o}dinger}
\newcommand{\Comm}[2]%
   {\ensuremath{[ \, #1, #2 \, ]}}
\newcommand{\MEangle}[3]%
   {\ensuremath{\langle \, #1 \, | \, #2 \, | \, #3 \, \rangle}}
\usepackage{hyperref}
\hypersetup{
    unicode=false,          
    pdftoolbar=true,        
    pdfmenubar=true,        
    pdffitwindow=false,     
    pdfstartview={FitH},    
    pdftitle={Paper},       
    pdfauthor={John Dawson},     
    pdfsubject={Phys Rev article},   
    pdfcreator={John Dawson},   
    pdfproducer={John Dawson}, 
    pdfkeywords={keyword1} {key2} {key3}, 
    pdfnewwindow=true,      
    colorlinks=true,        
    linkcolor=red,          
    citecolor=blue,         
    filecolor=green,        
    urlcolor=cyan           
}
\begin{document}
%
%
\preprint{LA-UR-16-22361}
\title{Stability of exact solutions of the nonlinear \Schrodinger\ equation in an  external potential having supersymmetry and parity-time symmetry} 

\author{Fred Cooper} 
\email{cooper@santafe.edu}
\affiliation{The Santa Fe Institute, 1399 Hyde Park Road, Santa Fe, NM 87501, USA}
\affiliation{Theoretical Division,
   Los Alamos National Laboratory,
   Los Alamos, NM 87545}
\author{Avinash Khare}
\email{khare@physics.unipune.ac.in}
\affiliation{Physics Department, Savitribai Phule Pune University, Pune 411007, India}
\author{Andrew Comech}
\email{comech@math.tamu.edu}
\affiliation{Department of Mathematics, Texas A\&M University, 
             College Station, TX 77843, USA}
\affiliation{Institute for Information Transmission Problems, Moscow 101447, Russia}
\author{Bogdan Mihaila}
\email{bmihaila@nsf.gov}
\affiliation{National Science Foundation, Arlington, VA 22230, USA}
\affiliation{
   Los Alamos National Laboratory,
   Los Alamos, NM 87545, USA}
\author{John F. Dawson}
\email{john.dawson@unh.edu}
\affiliation{Department of Physics,
   University of New Hampshire,
   Durham, NH 03824, USA}   
\author{Avadh Saxena} 
\email{avadh@lanl.gov}
\affiliation{Theoretical Division,
   Los Alamos National Laboratory,
   Los Alamos, NM 87545}
\date{\today, \now \ EDT}
%
%
%
\begin{abstract}
We  discuss the stability properties of the solutions of the general 
nonlinear \Schrodinger\  equation (NLSE)  in 1+1 dimensions in an 
external potential derivable from a parity-time  (${\mathcal P}{\mathcal T}$) symmetric  superpotential $W(x)$ that  we considered earlier [Kevrekidis \etal\ Phys. Rev. E \textbf{92}, 042901 (2015)].  In particular we consider the nonlinear partial differential equation 
$  \{
   i \,
   \partial_t
   +  
   \partial_x^2
   - 
   V^{-}(x)
   +
   | \psi(x,t) |^{2\kappa}
   \} \, \psi(x,t) 
   = 
   0 \>, $
for arbitrary nonlinearity parameter~$\kappa$.  We study the bound state solutions when $V^{-}(x) = (1/4- b^2) \sech^2(x)$, which can be derived from two different superpotentials $W(x)$, one of which is complex and ${\mathcal P}{\mathcal T}$ symmetric.  Using  Derrick's theorem, as well as a time dependent variational approximation, we derive exact analytic results for the domain of stability of the trapped solution as a function of the  depth $b^2$ of the external potential.  We compare the regime of stability found from these analytic approaches with a numerical linear stability analysis using a variant of the Vakhitov-Kolokolov (V--K) stability criterion.  The numerical results of applying the V-K condition give the \emph{same} answer for the domain of stability as the analytic result obtained from applying Derrick's theorem.  Our main result is that for $\kappa>2$ a \emph{new} regime of stability for the exact solutions appears as long as $b > b_{\text{crit}}$, where $b_{\text{crit}}$ is a function of the nonlinearity parameter $\kappa$.  In the absence of the potential the related solitary wave solutions of the NLSE
are \emph{unstable} for $\kappa>2$.
\end{abstract}
\maketitle

\section{\label{s:Intro}Introduction}

The topic of Parity-Time ($\mc{PT}$) symmetry and its relevance for physical 
applications on the one hand, as well as its mathematical structure
on the other, have drawn considerable attention from both the physics
and the mathematics community. Originally the proposal of Bender and his 
collaborators \cite{Bender_review,special-issues} towards the study of 
such systems was made as an alternative to the postulate of Hermiticity in 
quantum mechanics. In view of the formal similarity of the \Schrodinger\ equation with 
Maxwell's equations in the paraxial approximation, it was realized that such $\mc{PT}$ invariant systems can in fact be experimentally realized in optics \cite{review,Muga,PT_periodic,experiment}.  Subsequently, these efforts motivated experiments in several other areas including $\mc{PT}$ invariant electronic circuits \cite{tsampikos_recent,tsampikos_review}, mechanical circuits \cite{pt_mech}, and whispering-gallery microcavities \cite{pt_whisper}.

Concurrently, the notion of supersymmetry (SUSY) originally espoused in 
high-energy physics has also been realized in optics \cite{heinr1}.  The key 
idea is that from a given potential one can obtain a SUSY partner potential 
with both potentials possessing the same spectrum (with exception of possibly one eigenvalue) \cite{Gendenshtein,susy}.  An interplay of SUSY with $\mc{PT}$ symmetry is expected to be quite rich and is indeed useful in achieving transparent as well as one-way reflectionless complex optical potentials \cite{bagchi,ahmed,midya}.  

In a previous paper \cite{pt1} we explored the interplay between $\mc{PT}$ symmetry, SUSY and nonlinearity. 
In Ref.~\cite{pt1} we derived the exact solutions of the general nonlinear \Schrodinger\ equation (NLSE) with arbitrary nonlinearity in 1+1 dimensions when in an external potential given by a shape invariant \cite{Gendenshtein,avinashfred} supersymmetric and $\mc{PT}$ symmetric complex potential.  In particular, we considered the nonlinear partial differential equation
\begin{equation}\label{eqn1}
   \qty{
   i \,
   \partial_t
   +  
   \partial_x^2
   - 
   V^\pm(x)
   +
   | \psi(x,t) |^{2\kappa}
   } \, \psi(x,t) 
   = 
   0 \>,
\end{equation}
for arbitrary nonlinearity parameter $\kappa$, with
\begin{equation}\label{eqn1x}
   V^{\pm}(x)
   =
   W_1^2(x)\mp W_1'(x) \>,
\end{equation}
and the partner potentials arise from the superpotential 
\begin{equation}\label{eqn1y}
   W_1(x)
   =
   \qty( m - 1/2 ) \,  \tanh{x} 
   - 
   i b \, \sech{x} \>,
\end{equation}
giving rise to
\begin{subequations}\label{VpVm}
\begin{align}
   V^{+}(x)
   &= 
   \qty( -b^2 - m^2 + 1/4 ) \, \sech^2(x) 
   \label{eqn7a} \\
   & \quad
   - 
   2 i \, m \, b \, \sech(x) \, \tanh(x),
   \notag \\
   V^{-}(x)
   &=
   \qty( - b^2 - (m-1)^2 + 1/4 ) \, \sech^2(x)
   \label{eqn8a} \\
   & \quad
   -
   2 i \, \qty( m - 1 ) \, b \, \sech(x) \, \tanh(x) \>.
   \notag
\end{align}
\end{subequations}
For $m=1$, the \emph{complex} potential $V^{+}(x)$ has the same spectrum, apart from the ground state, as the \emph{real} potential $V^{-}(x)$ and we used this fact in our numerical study of the stability of the bound state solutions of the NLSE in the presence of $V^{+}(x)$ (see Ref.~\cite{pt1}).
To complete our study of this system of nonlinear \Schrodinger\ equations in $\mc{PT}$ symmetric SUSY external potentials, we will study the stability properties of the bound state solutions of NLSE in the presence of the external real SUSY partner potential $V^{-}(x)$, and compare the stability regime of these solutions, which depends on the parameters $(b, \kappa)$, to the stability regime of the related  \emph{solitary} wave solutions to the NLSE in the absence of the external potential, which depend only on the parameter $\kappa$.  Because the NLSE in the presence of $V^{-}(x)$ is a Hamiltonian dynamical system, we can use variational methods to study the stability of the solutions when they undergo certain small deformations.  We will compare the results of this type of analysis with a linear stability analysis based on the V--K stability criterion \cite{comech,stab1}.

In our previous paper \cite{pt1}  we determined the exact solutions of the equation for $m=1$ and for $V^{+}(x)$, which was complex. We studied numerically the stability properties of these solutions using linear stability analysis.  We found some unusual results for the stability which depended on the value of $b$.  In that paper, because of the complexity of the potential, the energy was not conserved and a Hamiltonian formulation of the problem was not possible.  However for the partner potential $V^{-}(x)$, when $m=1$, the potential is real.  We note that $V^{-}(x)$ has the symmetry $b \leftrightarrow m-1$, so that  we can obtain $V^{-}(x)$ from two \emph {different} superpotentials $W_1(x)$ and $W_2(x) $ that have the  ${\mathcal P}{\mathcal T}$ symmetric forms at arbitrary $m$: 
\begin{subequations}\label{SSW1W2}
\begin{align}
   W_1(x)
   &=
   \qty( m - 1/2 ) \, \tanh(x)
   - 
   i b \, \sech(x)  \>,
   \label{SSW1} \\
   W_2(x) 
   &= 
   ( b + 1/2 ) \, \tanh(x) 
   - 
   i (m-1) \sech(x) \>.
   \label{SSW2}
\end{align}
\end{subequations}
In particular, when $m=1$ we can determine  the spectrum of bound states of the linear \Schrodinger\ equation with potential $V^{-}(x)$ using the \emph{real} superpotential $W_2(x) = \qty( b + 1/2 ) \, \tanh{x}$ and the real shape invariant sequence of partner potentials.  When $m=1$  the first superpotential $W_1(x)$ has the  \emph{complex} partner potential $V^{+}(x)$ which we studied in \cite{pt1} , whereas $W_2(x)$   is the usual \emph{real}  shape invariant SUSY potential. Both superpotentials yield the same real potential  $ V^{-}(x) = -(b^2 - 1/4) \sech^2(x)$.  Because this potential is real, one can use variational methods to study the stability of the exact solutions to the NLSE in the potential $V^{-}(x)$.  We will consider both Derrick's theorem \cite{ref:derrick} as well as a time dependent variational approximation \cite{var1,var2} to study the stability of the exact solutions.  Because of the similarity of the solutions to those of the NLSE equation, we are able to use a variant of the V--K stability criterion to study spectral stability of the solutions.  The results of the V--K analysis agree with the new regime of stability found from Derrick's theorem and the time dependent variational approach. The latter approach allows us to obtain the frequency of small oscillations of the perturbed solutions.  We are able to obtain exact analytic results because we can formulate the problem in terms of Hamilton's action principle.  The Euler-Lagrange equations lead to the dynamical equations which have a conserved Hamiltonian.  This is in sharp contrast with the stability analysis for the solutions in the presence of $V^{+}(x)$ which had to be done numerically.  The latter system is dissipative in nature. 

This paper is structured as follows.  In Sec.~\ref{s:SSModel} we review the non-hermitian SUSY model that we studied in Ref.~\cite{pt1}.  In Sec.~\ref{s:HamPrinc} we consider Hamilton's principle for the NLSE in the real external potential $V^{-}(x)$.  Section~\ref{s:Derrick} describes the application of Derrick's theorem for determining the domain of stability of the solutions.  Here we determine analytically a new domain of stability for the solutions, when compared to the solutions in the absence of an external potential.  We find that for $b > b_{\text{crit}}$ a new domain of stability exists.  Sec.~\ref{s:collective} contains a collective coordinate approach which allows one to study dynamically the blowup or collapse of the solution as well as small oscillations around the exact solution when slightly perturbed.  By setting the frequency to zero we obtain analytically a domain of stability that agrees with the result of Derrick's theorem.  In Sec.~\ref{NS.s:LinearStability} we provide the details of a linear stability analysis based on the V-K stability criterion, which leads to identical conclusions as Derrick's theorem.  Section~\ref{s:conclude} contains a summary of our main results. 

%
%
\section{\label{s:SSModel}A Linear Non-Hermitian Supersymmetric Model}

We are motivated by the $\mathcal{PT}$ symmetric SUSY superpotential
\begin{equation}\label{eqn6}
   W_1(x)
   =
   \qty( m - 1/2 ) \,  \tanh(x) 
   - 
   i b \, \sech(x) \>,
\end{equation}
which gives rise to the supersymmetric partner potentials given by Eqs.~\ef{VpVm}.
In what follows we will specialize to the case $m=1$. In that case, $V^{-}(x)$  is the well known P{\"o}schl-Teller potential~\cite{poschl,LL}.  The relevant bound state eigenvalues assume an extremely simple form as  
\begin{equation}\label{eqn9}
   E_n^{(-)}
   =
   -\frac{1}{4} \, \qty[ \, 2 b - 2 n - 1 \, ]^2 \>.
\end{equation}
Such bound state eigenvalues only exist when $n < b - 1/2$.  We notice that for  the ground state (n=0) to exist requires $b > 1/2$.  The existence of a first excited state (n=1) requires $b > 3/2$.
We will find that the stability of the NLSE solutions in this external potential will depend on the depth of the well, which will lead to a critical value of $b = b_{\text{crit}}$, above which the solutions are stable.



In what follows we will be concerned with the properties of the NLSE in the presence of the external potential centered at $x=0$,
\begin{equation}\label{Vmdef}
   V^{-}(x) 
   = - \qty( b^2 - 1/4 ) \, \sech^2(x) \qc b > 1/2 \>.
\end{equation}
In particular we are interested in the bound state solutions of
\begin{equation}\label{eqn1-II}
   \qty{
   i \,
   \partial_t
   +  
   \partial_x^2
   - 
   V^{-}(x)
   +
   | \psi(x,t) |^{2\kappa}
   } \, \psi(x,t) 
   = 
   0 \> .
\end{equation}
If we assume a solution of the form 
\begin{equation}\label{eqn-form}
   \psi(x,t)
   =
   A \sech^\alpha(x) \, \exp{-i \omega t} \>,
\end{equation}
it is easy to show that an exact solution is given by:
\begin{equation}\label{exact}
   \psi(x,t) 
   = 
   A(b,\kappa) \, \sech^{1/\kappa}(x) \, e^{-i \omega t} \>,
\end{equation}
where $\omega = - 1/\kappa^2$ and   
\begin{equation}\label{Abound}
   A^{2\kappa}(b,\kappa) 
   = 
   ( 1/\kappa + 1/2)^2 - b^2 \>.
\end{equation}
For the complex potential $V^{+}(x)$, the amplitude of the solution is given by \cite{pt1}
\begin{equation*}
   A^{2\kappa}(b,\kappa)
   =
   \frac{ \qty( ( 1/\kappa + 1/2)^2 - m^2 )
          \qty( ( 1/\kappa + 1/2)^2 - b^2 ) }
        { \qty( 1/\kappa + 1/2 )^2 } \>,
\end{equation*}
so that there are two separate regimes where $A$ is real.  In contrast, for $V^{-}(x)$ 
there is only one regime for an attractive $V^{-}(x)$ where $A$ is real, 
namely
\begin{equation}\label{brange}
   b_{\text{min}} = 1/2 \leq b \leq b_{\text{max}} = (1/\kappa + 1/2) \>.
\end{equation}
The analysis of the stability of the solutions for $V^{+}(x)$ in Ref.~\cite{pt1} showed a very complicated pattern.  Even for $\kappa=1$ there is a regime of instability as a function of $b$ for the nodeless solution.  For $\kappa=3$ all solutions in that case found analytically and numerically were unstable. Only for $\kappa < 2/3$ were the solutions stable.  In contrast to that analysis where each  value of $\kappa$ had to be investigated separately, in the case of the real potential $V^{-}(x)$, we are able to address the stability question for all $\kappa$ using Derrick's theorem as well as the V--K stability criterion.  

The mass $M$ of the bound state for the case $V^{-}(x)$ is given by
\begin{align}
   M
   &=
   \int |\psi(x,t)|^2 \dd{x}
   \label{Mbound} \\
   &=
   \frac{ \sqrt{\pi} \, \qty[ (1/2 + 1/\kappa)^2 - b^2 ]^{1/\kappa} \, \Gamma[1/\kappa] }
        { \Gamma[1/2 + 1/\kappa] } \>.
   \notag
\end{align}
If we turn off the external potential by setting $b \to 1/2$, $A^{\kappa}[1/2,\kappa] \to (1/\kappa + 1/2)$ and the mass $M$ of the bound state goes to the mass of the solitary wave solutions,
\begin{equation}\label{Msol}
   M
   \to
   \frac{ \sqrt{\pi} \, \qty[ (\kappa + 1)/\kappa^2 ]^{1/\kappa} \, \Gamma[1/\kappa] }
        { \Gamma[1/2 + 1/\kappa] } \>.   
\end{equation}

%
%
\section{\label{s:HamPrinc}Hamilton's Principle of Least Action for the NLSE in an external potential} 

Let us first discuss Hamilton's principle of least action for the usual NLSE without a confining potential.
The NLSE with arbitrary nonlinearity in 1+1 dimensions is given by 
\begin{equation}\label{ham.e:1}
   \qty{
   i \,
   \partial_t
   +  
   \partial_x^2
   +
   g \, | \psi(x,t) |^{2\kappa}
   } \, \psi(x,t) 
   = 
   0 \>.   
\end{equation}
The second term causes diffusion  and the third term attraction and the competition
allows for solitary wave blowup which depends on $\kappa$.  Here $g$ can be scaled out of the equation by letting
\begin{equation}
   \psi(x,t) \mapsto  g^{-1/( 2\kappa)}\, \psi(x,t) \>,
\end{equation}
so that the linear equation for the rescaled equation is obtained in the 
limit $\psi(x,t) \rightarrow 0$.  While the solitary waves are stable for 
$\kappa < 2$, for $\kappa = 2$ there is a critical mass $M$ necessary for blowup to occur, where the width of solitary wave goes to zero.  For $\kappa > 2$, blowup occurs in a finite amount of time.  The classical action for the NLSE is $\Gamma[\psi,\psi^{\ast}] = \int L[\psi,\psi^{\ast}] \dd{t}$, where the Lagrangian $L[\psi,\psi^{\ast}]$ is given by
\begin{align}
   L[\psi,\psi^{\ast}]
   &=
   \int \dd{x}
   \bigl \{ \,
      i \, \qty[ \psi^{\ast} (\partial_t \psi) - (\partial_t \psi^{\ast} ) \psi ] / 2
      \label{LagNLSE} \\
      & \qquad
      +   
      ( \partial_x \psi^{\ast} )( \partial_x \psi)
      -
      | \psi |^{2(\kappa+1)} / ( \kappa + 1 ) \,
   \bigr \} \>.
   \notag
\end{align}
The NLSE follows from the Hamilton's principle of least action, $\delta \Gamma / \delta \psi = 0$ and $\delta \Gamma / \delta \psi^{\ast} = 0$, which leads to Eq.~\ef{ham.e:1} with $g \to 1$.  Multiplying this equation by $\psi^{\ast}(x,t)$ and subtracting its complex conjugate, it is easy to prove that the mass $M$, defined by $M = \int |\psi(x,t)|^2 \dd{x}$, is conserved.  We now want to add a \emph{real} SUSY potential to the NLSE.  We will consider the addition of $V^{-}(x)$ given in Eq.~\ef{Vmdef} so that the equation of motion is now given by
\begin{equation}\label{motion2}
   \qty{
   i \,
   \partial_t
   +  
   \partial_x^2
   - 
   V^{-}(x)
   +
   | \psi(x,t) |^{2\kappa}
   } \, \psi(x,t) 
   = 
   0 \>.  
\end{equation}
The action which leads to Eq.~\ef{motion2} is given by $\Gamma[\psi,\psi^{\ast}] = \int L[\psi,\psi^{\ast}] \dd{t}$ where
\begin{align}
   L[\psi,\psi^{\ast}]
   &=
   \int \dd{x} i \, \qty[ \psi^{\ast} (\partial_t \psi) - (\partial_t \psi^{\ast} ) \psi ]/2
   -
   H[\psi,\psi^{\ast}] \>,
   \notag \\
   H[\psi,\psi^{\ast}]
   &=
   \int \dd{x} 
   \bigl \{ \,
       ( \partial_x \psi^{\ast} )( \partial_x \psi)
       -
       | \psi |^{2(\kappa+1)} / (\kappa + 1)
       \notag \\
       & \qquad\qquad
       +
       \psi^{\ast} \, V^{-}(x) \, \psi \,
    \bigr \} \>.
    \label{Hdef}
\end{align}

%
%
\section{\label{s:Derrick}Derrick's theorem} 

Derrick's theorem \cite{ref:derrick} states that for a Hamiltonian dynamical system, for a solitary wave solution to be stable it must be stable to changes in scale transformation $x \mapsto \beta x$ when we keep the mass of the solitary wave fixed.  That is the Hamiltonian needs to be a minimum in $\beta$ space.  First let us look at the case of the NLSE without an external potential: Derrick's method is based on whether a scale transformation which keeps the mass $M$ invariant, raises or lowers the energy of a solitary wave. For the NLSE with Hamiltonian 
\begin{align}
   H 
   &=
   \int \dd{x}  
   \qty[ 
      ( \partial_x \psi^\ast ) \, ( \partial_x \psi )
      - 
      | \psi |^{2(\kappa+1)} / (\kappa + 1) ]
   \\
   & 
   \equiv 
   H_1 - H_2 \>,
   \notag
\end{align}
where both $H_1$ and~$H_2$ are positive definite.  A static solitary wave solution can be written as
\begin{equation}
   \psi(x,t) 
   = 
   r(x) \, e^{- i \omega t} \>.
\end{equation}
The exact solution has the property that it minimizes the Hamiltonian subject 
to the constraint of fixed mass as a function of a stretching factor $\beta$.
This can be seen by studying a variational approach as done in  ~\cite{variational},
or by directly studying the effect of a scale transformation that respects conservation of mass. 
In the latter approach, which generalizes the method used by Derrick ~\cite{ref:derrick}, we let $x \mapsto \beta x$, and consider the stretched wave function,
\begin{equation}\label{stretchpsi}
   \psi_{\beta}(x,t) 
   = 
   \beta^{1/2} r(\beta x) \, e^{ - i \omega t} \>,
\end{equation}
so that
\begin{equation*}
   M 
   =
   \int \dd{x} \abs{ \psi_\beta(x,t) }^2
   = 
   \int \dd{x} \abs{ \psi(x,t) }^2
\end{equation*}
is preserved by the transformation.  Defining $H_\beta$ as the value of $H$ for the stretched solution $\psi_\beta(x,t)$, one finds that $\partial H_{\beta} / \partial \beta |_{\beta=1} = 0$ is consistent with the equations of motion.  The stable solutions must then also satisfy:
\begin{equation}\label{d2Hbeta-dbeta2}
   \pdv[2]{H_{\beta}}{\beta} \ge 0 \>.
\end{equation}
If we write $H$ in terms of the two positive definite pieces $H_1$, $H_2$, then
\begin{equation}\label{H1H2defs}
   H_{\beta}
   =
   \beta^2 \, H_1 - \beta^{\kappa} \, H_2 \>,
\end{equation}
we find
\begin{equation}\label{H1andH2}
   \eval{ \pdv{H_{\beta}}{\beta} }_{\beta=1}
   =
   2 \, H_1 - \kappa \, H_2
   =
   0 \>,
\end{equation}
so that $H_1 = (\kappa / 2 ) \, H_2$.  This result is consistent with the equations of motion.
In fact for the NLSE the exact solution has $r(x) = A \, \sech^{1/\kappa}(x)$ where $A^{2\kappa}= (\kappa + 1)/\kappa^2$.  One finds then using
\begin{equation*}
   \int_{-\infty}^\infty \dd{x}
   \sech^r(x) 
   = 
   \frac{\sqrt{\pi} \, \, \Gamma[ r/2] }{\Gamma[1/2+r/2]} \>,
\end{equation*}
that
\begin{subequations}\label{H1andH2eval}
\begin{align}
   H_1
   &=
   \frac{ \sqrt{\pi} \, \qty[ (\kappa + 1)/\kappa^2 ]^{1/\kappa} \Gamma[1/\kappa] }
        { 2 \kappa^2 \, \Gamma[3/2 + 1/\kappa] } \>,
   \label{H1-val} \\
   H_2
   &=
   \frac{ \sqrt{\pi} \, \qty[ (\kappa + 1)/\kappa^2 ]^{1/\kappa} \Gamma[1/\kappa] }
        { \kappa^3 \, \Gamma[3/2 + 1/\kappa] } \>,
   \label{H2-val}   
\end{align}
\end{subequations}
so that the exact solution is indeed a minimum of the Hamiltonian with respect to scale transformations, with $H_1 = (\kappa/2) \, H_2$.

The second derivative is given by 
\begin{equation}\label{d2Hbeta}
   \pdv[2]{H_{\beta}}{\beta}
   =
   2 \, H_1 - \kappa(\kappa - 1) \, \beta^{\kappa - 2} \, H_2 \>,
\end{equation}
which when evaluated at the stationary point yields
\begin{equation}\label{d2H2stable}
   \pdv[2]{H_{\beta}}{\beta}
   =
   2 \, (2 - \kappa) \, H_1 \ge 0 \>,
\end{equation}
for stability.  This result indicates that solutions are unstable to changes in the 
width, compatible with the conserved mass, when $\kappa > 2$.  The case $\kappa=2$ is a marginal case where it is known that blowup occurs at a critical mass (see for example Ref.~\cite{var2}).  The result found above for the NLSE has also been found by various other methods such as linear stability analysis and using strict inequalities.  Numerical simulations (see Ref.~\cite{Rose:p}) have been done for the critical case $\kappa=2$ showing that blowup (self-focusing) occurs when the mass $M > 2.72$.  For $\kappa >2$ a variety of analytic and numerical methods have been used to study the nature of the blowup at finite time~\cite{kevrekedis}.  

%
%
\subsection{\label{s:VKcriterion}Linear Stability and the Vakhitov--Kolokokov criterion}

In the case of the nonlinear \Schrodinger\ equation, one can perform a linear stability analysis of the exact solutions.  Namely one lets
\begin{equation}\label{nlse.e:1}
   \psi(x,t) 
   = 
   \qty[ \psi_\omega(x) + r(x,t) ] \, e^{-i \omega t} \>,
\end{equation}
and linearizes the NLSE to find an equation for $r(x,t)$ to first order,
\begin{equation}\label{nlse.e:2}
   \partial_t \, r(x,t)
   =
   A_{\omega} \, r(x,t) \>,
\end{equation}
and studies the eigenvalues of the differential operator $A_\omega$.  If the spectrum of $A_\omega$ is imaginary, then the solutions are spectrally stable.  V--K showed \cite{stab1} that when the spectrum is purely imaginary $\dd M(\omega)/\dd \omega < 0$.  Also they showed that when $\dd M(\omega)/\dd \omega > 0$, there is a real positive eigenvalue so that there is a linear instability.
For the NLSE, there is a class of solutions with arbitrary nonlinearity parameter $\kappa$.  Namely
\begin{subequations}\label{nlse.e:3}
\begin{align}
   \psi_{\omega}(x,t)
   &=
   A(\kappa,\beta) \, \sech^{1/\kappa}(\beta x) \, e^{-i \omega t} \>,
   \label{nlse.e:3-a} \\
   A^{2\kappa}(\kappa,\beta)
   &=
   \beta^2 ( \kappa + 1 ) /\kappa^2
   \qc
   \omega = - \beta^2 / \kappa^2 \>.
   \label{nlse.e:3-b}
\end{align}
\end{subequations}
When we do not have an external potential, we know explicitly how the mass changes when we change $\omega$ at fixed $\kappa$.  That is
\begin{align}
   M(\omega)
   &=
   \int_{-\infty}^{+\infty} \!\!\! \dd{x} \abs{ \psi_{\omega}(x,t) }^2
   =
   A^2(\beta,\kappa) \, C_1(\kappa) / \beta 
   \label{nlse.e:4} \\
   &=
   \frac{ \sqrt{\pi} \, \qty[ - \omega / (\kappa + 1) ]^{1/\kappa} \Gamma[1/\kappa] }
        { \kappa \, \sqrt{-\omega} \, \Gamma[1/2 + 1/\kappa] } \>,
   \notag      
\end{align}
where
\begin{equation}\label{nlse.e:5}
   C_1(\kappa)
   =
   \int_{-\infty}^{+\infty} \!\!\! \dd{x} \sech^{2/\kappa}(x) \>.
\end{equation}
We find
\begin{equation}\label{nlse.e:6}
   \dv{M}{\omega}
   =
   a_1 \, (\kappa - 2) 
   \qc
   a_1 > 0 \>.
\end{equation}
Thus for $\kappa > 2$ the solitary waves are unstable.  This agrees with the result of Derrick's theorem.  When we have an external potential, we will need to determine the solutions numerically as we change $\omega$.  This will be accomplished in Sec.~\ref{NS.s:LinearStability}.

%
%
\subsection{\label{ss:Derrick-external}Adding an external potential} 

So now let us look at our situation when we have in addition the real external potential:
\begin{equation}\label{Dext.e:1}
   V^{-}(x)
   = 
   - (b^2 - 1/4) \sech ^2(x) \>.
\end{equation}
The exact solution to the NLSE in the presence of $V^{-}(x)$ is given by Eq.~\ef{exact}.
This  solution is  similar in form to the usual solution to the  NLSE except this nodeless solution is pinned to the potential so that there is no translational invariance. When $b=1/2$ this solution goes over to a particular solution of the NLSE with width parameter $\beta=1$.  Under the scale transformation $x \to \beta x$, the stretched solution which preserves the mass $M$ is given by:
\begin{subequations}\label{Dext.e:2}
\begin{align}
   \psi_{\beta}(x,t)
   &=
   A(b,\kappa) \, \beta^{1/2} \, \sech^{1/\kappa}(\beta x) \, e^{-i \omega t} \>,
   \label{Dext.e:2-a} \\
   A^{2\kappa}(b,\kappa)
   &=
   \qty( 1/2 + 1/\kappa )^2 - b^2
   \qc
   \omega = - 1 / \kappa^2 \>.
   \label{Dext.e:2-b}
\end{align}
\end{subequations}
The stretched wave function $\psi_{\beta}(x,t)$ is no longer an exact solution.
The stretched Hamiltonian for the external potential case is now given by
\begin{equation}\label{Dext.e:3}
   H_{\beta} 
   = 
   \beta^2 H_1(b,\kappa) 
   - 
   \beta^{\kappa} \, H_2(b,\kappa) 
   + 
   H_3(b,\kappa,\beta) \>,
\end{equation}
where
\begin{align*}
   H_1(b,\kappa)
   &=
   A^2(b,\kappa) \, f_1(\kappa) , 
   \\
   f_1(\kappa)
   &=
   \frac{ \sqrt{\pi} \, \Gamma[1/\kappa] }
        { 2 \kappa^2 \, \Gamma[3/2+1/\kappa] } \>,      
\end{align*}
and
\begin{align*}
   H_2(b,\kappa)
   &=
   A^{2(\kappa+1)}(b,\kappa) \, f_2(\kappa) , 
   \\
   f_2(\kappa)
   &=
   \frac{ \sqrt{\pi} \, \Gamma[1+1/\kappa] }
        { (\kappa + 1) \, \Gamma[3/2+1/\kappa] } \>,    
\end{align*}
with $A(b,\kappa)$ now given by \ef{Dext.e:2-b}, and
\begin{align*}
   &H_3(b,\kappa,\beta)
   =
   \int_{-\infty}^{+\infty} \!\!\!\!
   \psi^{\ast}_{\beta}(x,t) \, V^{-}(x) \, \psi_{\beta}(x,t) \dd{x}
   \\
   &
   =
   \qty(1/4 - b^2) \, A^2(b,\kappa) \, 
   \int_{-\infty}^{+\infty} \!\!\!\!\! 
   \beta \, \sech^{2/\kappa}(\beta x) \, \sech^2(x) \dd{x} \>.
\end{align*}
Thus, we find
\begin{align*}
   &\eval{ \pdv{H_3}{\beta} }_{\beta=1}
   \!\!\!\!
   =
   \qty(1/4 - b^2) \, A^2(b,\kappa)
   \\
   &
   \! \times
   \int_{-\infty}^{+\infty} \!\!
   [\, 
      \sech^{2/\kappa + 2}(x)
      -
      2 x \, \sech(x) \, \sech^{2/\kappa + 3}(x) / \kappa \,] 
      \dd{x} \>.
\end{align*}
Using the identity,
\begin{align}
   &\pdv{\sech^{2+2/\kappa}( \lambda x )}{\lambda}
   \label{Dext.e:4} \\
   & \quad
   =
   -
   (2/\kappa + 2) \, x \, \sinh(\lambda x) \, \sech^{2/\kappa + 3}( \lambda x) \>,
   \notag
\end{align}
we obtain
\begin{equation}\label{Dext.e:5}
   \eval{ \pdv{H_3}{\beta} }_{\beta=1}
   \!\!\! =
   -
   A^{2}(b,\kappa) \,
   \frac{ \sqrt{\pi} \, (1/4 - b^2 ) \, \Gamma[1/\kappa] }
        { (\kappa + 1) \, \Gamma[3/2+1/\kappa] } \>.   
\end{equation}
As in the case when $V^{-}(x) = 0$, we again find
\begin{equation}\label{Dext.e:6}
   \eval{ \pdv{H_{\beta}}{\beta} }_{\beta=1}
   \!\!\! =
   0   
\end{equation}
for our exact solution.  So the stretched solution is again an extremum of $H_\beta$ with $M$ kept fixed.

\begin{figure}[t]
   \centering
   \subfigure[\ $d^2 H/d \beta^2$ vs $b$ for $\kappa = 2.1$.]
   { \label{NS.f:dHdbeta}
     \includegraphics[width=0.85\columnwidth]{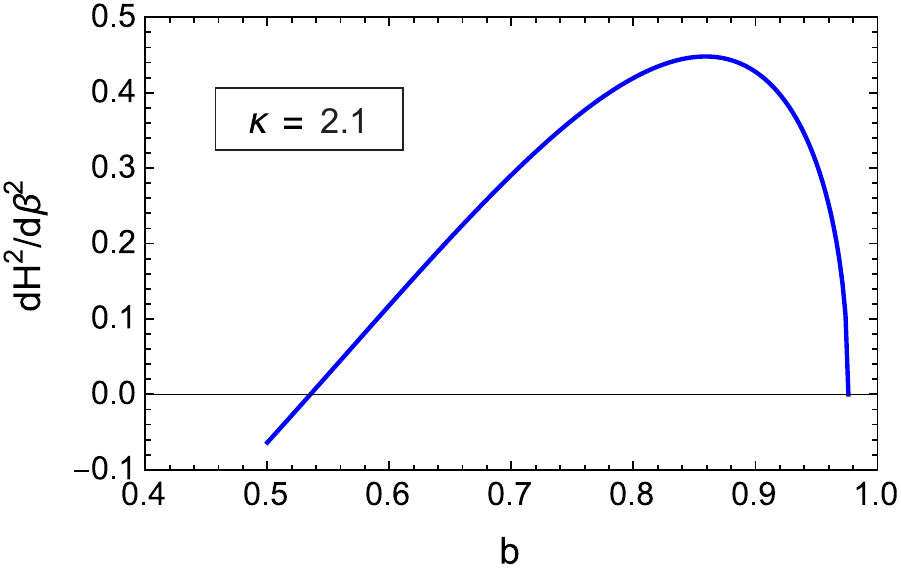} }
   \subfigure[\ $b$ vs $\kappa$.]
   { \label{NS.f:bvskappa}
     \includegraphics[width=0.85\columnwidth]{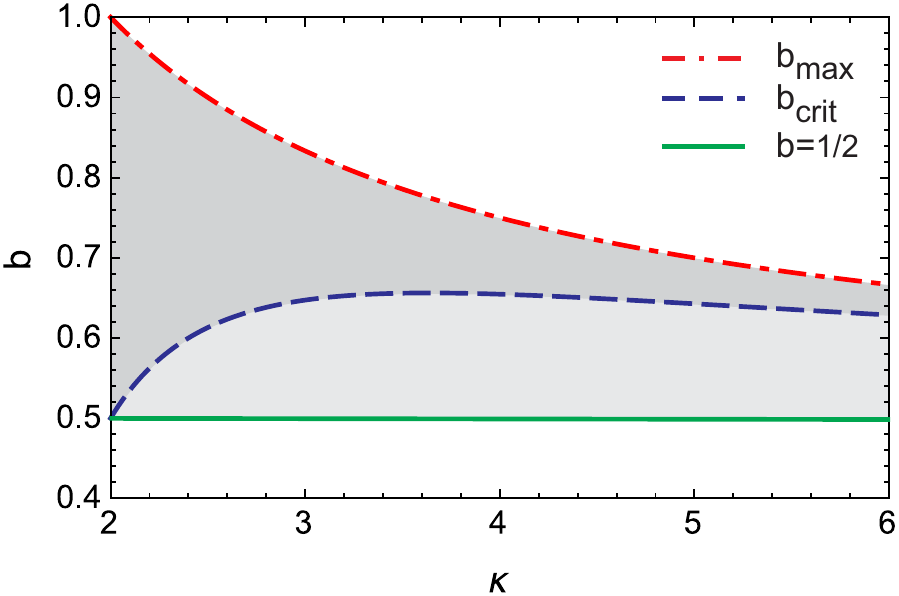} }
   \subfigure[\ $M$ vs $\kappa$.]
   { \label{NS.f:Mcrit}
     \includegraphics[width=0.85\columnwidth]{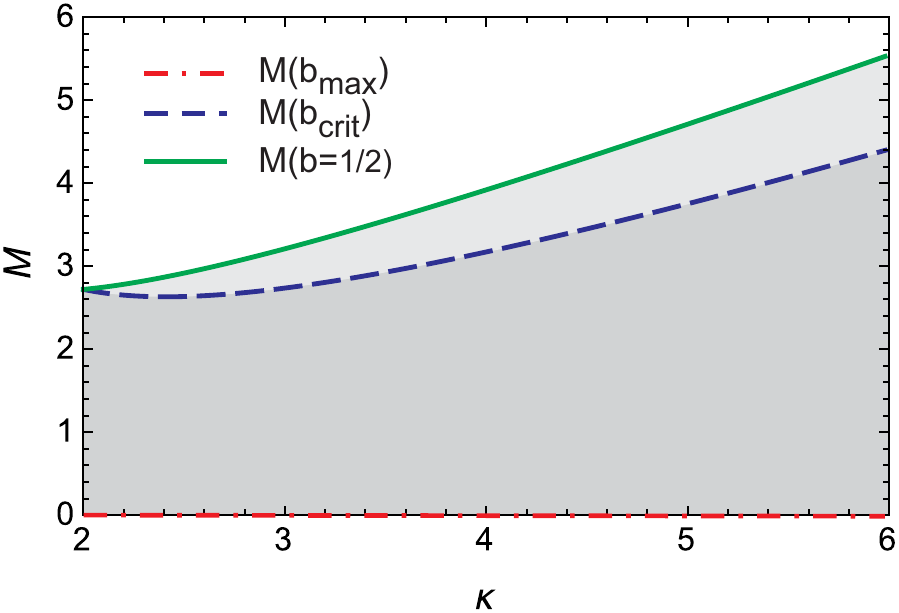} }
   \caption{\label{NS.f:Derrick}(a) $d^2 H/d \beta^2$ vs $b$ for $\kappa = 2.1$. (b) Here we plot
     $b_{crit}$ and $b_{max} $  vs $\kappa$.  The darker shaded area  is the predicted region of 
       stability for $\kappa > 2$ and the lighter shaded region  is unstable. (c) Here we plot $M$ vs $\kappa$. The lightly shaded area  is unstable, and the darker region is stable.}
\end{figure}

For the second derivative we have
\begin{equation}\label{Dext.e:7}
   \eval{ \pdv[2]{H_{\beta}}{\beta} }_{\beta=1}
   \!\!\! =
   2 \, H_1 
   - 
   \kappa ( \kappa - 1 ) \eval{ \pdv[2]{H_{3}}{\beta} }_{\beta=1} \>,
\end{equation}
where
\begin{equation}\label{Dext.e:8}
   \eval{ \pdv[2]{H_{3}}{\beta} }_{\beta=1}
   \!\!\! =
   (1/4 - b^2 ) \, A^{2}(b,\kappa) \,
   \qty[ I_1 + I_2 + I_3 ] / \kappa \>,
\end{equation}
with
\begin{align*}
   I_1
   &=
   - \int \dd{x}
   2 x^2 \, \sech^{2/\kappa + 2}(x) \>,
   \\
   I_2
   &=
   \int \dd{x}
   2 ( 2/\kappa + 1 ) \, x^2 \, \sinh^2(x) \, \sech^{2/\kappa + 4}(x) \>,
   \\
   I_3
   &=
   - \int \dd{x}
   4 x \, \sinh(x) \, \sech^{2/\kappa + 3}(x) \>.
\end{align*}
We can again evaluate these integrals using the first identity Eq.~\ef{Dext.e:4} and the identity:
\begin{align}
   &\pdv[2]{\sech^{2+2/\kappa}( \lambda x )}{\lambda}
   \label{Dext.e:9} \\
   & \quad
   =
   (2/\kappa +2)(2/\kappa + 3) \,
   x^2 \sinh^2(\lambda x) \, \sech^{2/\kappa + 4}(\lambda x)
   \notag \\
   & \qquad\qquad
   -
   (2/\kappa + 2) \, x^2 \, \sech^{2/\kappa + 2}(\lambda x) \>.
   \notag
\end{align}
We will also need the following hypergeometric function:
\begin{align}
   &ug(\kappa)
   =
   \int \dd{x} x^2 \, \sech^{2/\kappa + 2}(x)
   =
   \frac{2^{(\kappa + 2)/\kappa} \kappa^3}{(\kappa+1)^3}
   \label{Dext.e:10} \\   
   & \times {}_4F_3( \,
         1+1/\kappa,
         1+1/\kappa,
         1+1/\kappa,
         2+2/\kappa; 
         \notag \\
         & \qquad\qquad
         2+1/\kappa,
         2+1/\kappa,
         2+1/\kappa;
         -1 \, ) \>.
   \notag
\end{align}
Using these results, Eq.~\ef{Dext.e:8} gives
\begin{align}
   \eval{ \pdv[2]{H_{3}}{\beta} }_{\beta=1}
   \!\!\!\!\! &=
   (1/4 - b^2 ) \, A^{2}(b,\kappa) \, f_3(\kappa) \>,
   \label{Dext.e:11} \\
   f_3(\kappa)
   &=
   -
   \qty[
      \frac{4 \, ug(\kappa)}{2+3\kappa} 
      +
      \frac{4 \sqrt{\pi} \, \kappa \, \Gamma[1+1/\kappa]}
           {(\kappa+1)(3\kappa+2) \, \Gamma[3/2+1/\kappa] } 
       ] .
   \notag
\end{align}
The critical value is determined from:
\begin{align}
   \eval{ \pdv[2]{H_{\beta}}{\beta} }_{\beta=1}
   \!\!\! 
   &=
   A^{2}(b,\kappa) \,
   \bigl [ \,
      2 \, f_1(\kappa)
      \label{Dext.e:12} \\
      & \hspace{-2em}
      -
      \kappa ( \kappa + 1 ) \, A^{2\kappa}(b,\kappa) \, f_2(\kappa)
      +
      ( 1/4 - b^2 ) \, f_3(\kappa) \,
      \bigr ] \>.
      \notag
\end{align}
Solving for the critical value of $b^2$, we find
\begin{equation}\label{Dext.e:13}
   b_{\text{crit}}^2 
   =
   \frac{ 
      \qty( \kappa^3 + 3 \kappa^2 - 8 \kappa - 4 ) \, f_2(\kappa) 
            - \kappa \, f_3(\kappa) }
        { 4 \kappa^2 ( \kappa  - 1 ) \, f_2(\kappa) - 4 \kappa \, f_3(\kappa) ) } . 
\end{equation}
The result of calculating the second derivative at $\beta=1$ and setting it equal to zero is that the domain of stability is now as follows: for $\kappa < 2$ and all $b$ in the range $1/2 < b < b_{\text{max}} = 1/\kappa + 1/2$, the solution is stable, as it was for the solitary wave solutions of the NLSE. Here $b=1/2$ corresponds to no external potential.  When $\kappa > 2$ the solitary wave solutions of the NLSE were \emph{unstable}. Instead, in the presence of the confining potential,  a new domain of stability  occurs when $\kappa > 2$ as long as  $b_{\text{crit}} < b < b_{max}$, where $b_{\text{crit}}$ is given by Eq.~\ef{Dext.e:13}.  We see this in the result for $\kappa = 2.1$ shown in Fig.~\ref{NS.f:dHdbeta}.  In Fig.~\ref{NS.f:bvskappa}, we show both $b_{\text{crit}}(\kappa)$ and $b_{\text{max}}(\kappa)$ as a function of $\kappa$.  The region between $b=1/2$ and $b_{\text{crit}}(\kappa)$ is unstable. As we will show in Sec.~\ref{NS.s:LinearStability}, this analytic result for $b_{\text{crit}}$ given by Eq.~\ef{Dext.e:13} is confirmed by our linear stability analysis. 

%
%
%
%
%
%

Just as there is a critical mass for instability in the NLSE at $\kappa=2$, for $ \kappa >2$
we can interpret the critical value of $b$ in terms of a critical mass which depends on $\kappa$  above which the solution is unstable.  Since from Eq. \eqref{Mbound}, we have 
\begin{equation}\label{PF.e:Mbound-II}
   M 
   = 
   \frac{ \sqrt{\pi} \, \qty[ (1/2 + 1/\kappa)^2 - b^2 ]^{1/\kappa} \, \Gamma[1/\kappa]}
        { \Gamma[1/2 + 1/\kappa] } \>,
\end{equation}
we see that the mass $M$ \emph{decreases} as we increase $b$ at fixed $\kappa$.  So as we go from the unstable case $b=1/2$ (no potential) and increase $b$ we decrease the mass until we reach an $M_{\text{crit}}$ below which the solution is stable. Finally we reach the curve $M=0$ which corresponds to $b=b_{\text{max}}[\kappa] = 1/2 + 1/\kappa$.  The different regimes are shown in Fig.~\ref{NS.f:Mcrit}.  In the lightly shaded regime the solutions are unstable.  The maximum value of the mass is given by the case $b=1/2$, when there is no longer a stabilizing potential.  The interval $b_{\text{crit}} < b < b_{\text{max}}$ corresponds to the regime $0 <  M < M_{\text{crit}}$.  This is the stable regime denoted by the darker shaded area in Fig.~\ref{NS.f:Mcrit}.


%
%
\section{\label{s:collective}Collective coordinate approach for studying perturbations to the exact solution}

In order to follow the time evolution of a slightly perturbed solitary wave or bound solution to a Hamiltonian dynamical system, without solving numerically the time dependent partial differential equations for $\psi(x,t)$, one can introduce time-dependent collective coordinates assuming that the general shape of the original solution is maintained apart from the height, width, and position, etc.  This will allow us to see whether these parameters just oscillate around the original values or whether the parameters grow or decrease in time.  When instabilities are seen in the variational results, it suggests that the exact solutions are also unstable.  Unlike Derrick's theorem when applied to the NLSE, the collective coordinate method can be applied to the special case  $\kappa=2$.  It also gives an approximate description of wave function blow-up or collapse in the unstable regime, and oscillation of the perturbed solution in the stable regime.  In the next section, we first apply this approach when there is no external potential.

%
%
\subsection{\label{ss:blowup}Self-similar analysis of blowup and critical mass for the NLSE}

Let us remind ourselves of the collective coordinate variational approach to 
blow-up for the NLSE \cite{var1,var2} with no external potential. 
Using this method, we found previously that when $\kappa = 2$, there is a critical value of the mass required before blowup could take place.  Derrick's theorem has nothing to say about the stability of the solitary wave solution for this case. 
To make the collective coordinate approach concrete, we assume self-similar solutions of the form: 
\begin{align}\label{var.e:1}
   \psi(x,t) 
   &= 
   A(t) \, f[ \beta(t) y(t) ] 
   \\
   & \qquad
   \times \exp[ i \qty( v \, y(t) / 2 + \Lambda(t) \, y^2(t) - \omega t ) ] \>.
   \notag
\end{align}
Here $\Lambda(t)$, $A(t)$, and $\beta(t)$ are arbitrary real functions of time alone, and $y(t) = x - q(t)$.  For no external potential translation invariance gives $q(t) = v_0 t$.  In particular at $t = 0$ and $v_0 = 0$, we will start with the exact solution of the form $\psi(x,0) = A \, \sech^{1/\kappa}(\beta x)$ and assume that this solution just changes during the time evolution in amplitude and width.  With this assumption one can derive the dynamical equations for $A(t)$ and $\beta(t)$ from Hamilton's principle of least action with the Lagrangian given in Eq.~\ef{LagNLSE}.  Noether's theorem yields three conservation laws: conservation of probability, conservation of momentum, and conservation of energy.  Conservation of probability gives ``mass'' conservation: 
\begin{equation}\label{var.e:2}
   M 
   = 
   \int_{-\infty}^{\infty } \!\!\! \dd{x} \abs{\psi(x,t)}^2 
   = 
   \frac{A^2(t)}{\beta(t)} \, \int_{-\infty}^{\infty }  \!\!\! \dd{z} f^2(z) \>, 
\end{equation}
and allows one to rewrite $A(t)$ in terms of the conserved mass $M$, the width  
parameter $\beta(t)$, and a constant $C_1$ whose value depends on $f(z)$.  Thus, 
\begin{equation}\label{var.e:3}
   A^2(t) = \frac {M \beta(t)}{C_1} \qc
   C_1 = \int_{-\infty}^{\infty } \!\!\! \dd{z} f^2(z) \>.
\end{equation}
We will therefore keep $M$ in our definition of $A(t)$ since it will be a relevant parameter when $\kappa=2$.  For $f(z) = \sech^{1/\kappa}(z)$, one obtains 
\begin{equation}\label{var.e:4}
   C_1
   = 
   \frac{\sqrt{\pi } \, \Gamma[ 1/\kappa ]}{\Gamma[1/2+1/\kappa]} \>.
\end{equation}
Setting $\beta(t) = 1/G(t)$, in terms of the new collective coordinates $\qty[G,\Lambda]$, the Lagrangian \ef{LagNLSE} is given by
\begin{equation}\label{var.e:5}
   L[G,\Lambda]
   =
   K[G,\Lambda] - H[G,\Lambda] \>,
\end{equation}
where
\begin{subequations}\label{var.e:6}
\begin{align}
   \frac{K[G,\Lambda]}{M}
   &=
   \frac{i}{2 M} \int \dd{x}  
   \qty[ \psi^{\ast} (\partial_t \psi) - (\partial_t \psi^{\ast} ) \psi ]
   \notag \\
   &=
   \frac{1}{2} \, v^2
   +
   \omega
   -
   \dot{\Lambda} \, G^2 \, \frac{C_2}{C_1} \>,
   \label{var.e:6a} \\
   \frac{H[G,\Lambda]}{M}
   &=
   -
   \frac{1}{M}
   \int \dd{x}
   \qty[ 
      ( \partial_x \psi^{\ast} )( \partial_x \psi)
      +
      | \psi |^{2(\kappa+1)} / ( \kappa + 1 ) ]
   \notag \\
   &  \hspace{-3em} =
   \frac{v^2}{4}
   +
   \frac{C_3}{C_1} \, \frac{1}{G^2}
   +
   4 \Lambda^2 \, \frac{C_2}{C_1} \, G^2
   -
   \frac{1}{(\kappa + 1)} \,  \frac{C_4}{C_1} \,
   \qty( \frac{M}{C_1 G} )^{\kappa} \>,
   \label{var.e:6b}
\end{align}
\end{subequations}
where
\begin{subequations}\label{var.e:7}
\begin{align}
   C_2
   &=
   \int_{-\infty}^{\infty } \!\!\! \dd{z} z^2 f^2(z)
   \label{var.e:7a} \\
   &=
   2^{(2/\kappa - 1)} \kappa^3 \, {}_4F_3( \,
         1/\kappa,
         1/\kappa,
         1/\kappa,
         2/\kappa; 
         \notag \\
         & \qquad\qquad
         1+1/\kappa,
         1+1/\kappa,
         1+1/\kappa;
         -1 \, ) \>,
   \notag \\
   C_3
   &=
   \int_{-\infty}^{\infty } \!\!\! \dd{z} \qty( f'(z) )^2
   \label{var.e:7b} \\
   &=
   \frac{\sqrt{\pi} \, 
   \Gamma[ 1+1/\kappa ]}{2 \kappa \, \Gamma[3/2+1/\kappa]} \>,
   \notag \\
   C_4
   &=
   \int_{-\infty}^{\infty } \!\!\! \dd{z} f^{(2 \kappa + 2)}(z)
   \label{var.e:7c} \\
   &=
   \frac{\sqrt{\pi} \, 
   \Gamma[ 1+1/\kappa ]}{2 \kappa \, \Gamma[3/2+1/\kappa]}
   =
   2 \kappa \, C_3 \>.
   \notag
\end{align}
\end{subequations}
Collecting terms from \ef{var.e:5} and \ef{var.e:6}, the Lagrangian is given by
\begin{align}
   \frac{L[G,\Lambda]}{M}
   &=
   \frac{1}{4} \, v^2
   +
   \omega
   -
   \dot{\Lambda} \, G^2 \, \frac{C_2}{C_1}
   -
   \frac{C_3}{C_1} \, \frac{1}{G^2}
   \label{var.e:8} \\
   & \qquad
   -
   4 \Lambda^2 \, \frac{C_2}{C_1} \, G^2
   +
   \frac{1}{(\kappa + 1)} \,  \frac{C_4}{C_1} \,
   \qty( \frac{M}{C_1 G} )^{\kappa} \>.   
   \notag
\end{align}
From the Euler-Lagrange equations we obtain the second order differential equation for $G$,
\begin{equation}\label{var.e:9}
   \ddot{G} 
   = 
   4 \, \frac{C_3}{C_2} \, \frac{1}{G^3}  
   - 
   \frac{4 \, \kappa^2}{(\kappa + 1)} \, 
   \frac{ C_3}{C_2 G} \, \qty( \frac{M}{C_1 G} )^{\kappa} \>,
\end{equation}
and the relation $\Lambda = \dot{G}/(2G)$.  
In solving these equations, we will use for the mass $M$, when we are not at the critical value $\kappa=2$, the expression for the mass for the solitary wave solution given by Eq.~\ef{Msol}.
If we do this, we can rewrite Eq.~\ef{var.e:9} as
\begin{equation}\label {gddot2}
 \ddot G = 4 \frac{C_3}{C_2}   \frac{1}{ G^3 }  - \frac{C_3}{C_2 } \frac{4}{G^{\kappa+1} }  \>.
\end{equation}
One notices that for $G[0]=1$, $\ddot G =0$, as it must for an exact solution.  We see that to get $G \rightarrow 0$ when $\kappa=2$ we need to have $M > M^{\star}$ where $M^{\star}$ is the value of the mass for the exact solution.  So initial conditions with a mass greater than this are necessary to see blow up at $\kappa=2$. 

\begin{figure}[t]
  \centering
  \includegraphics[width=0.9\columnwidth]{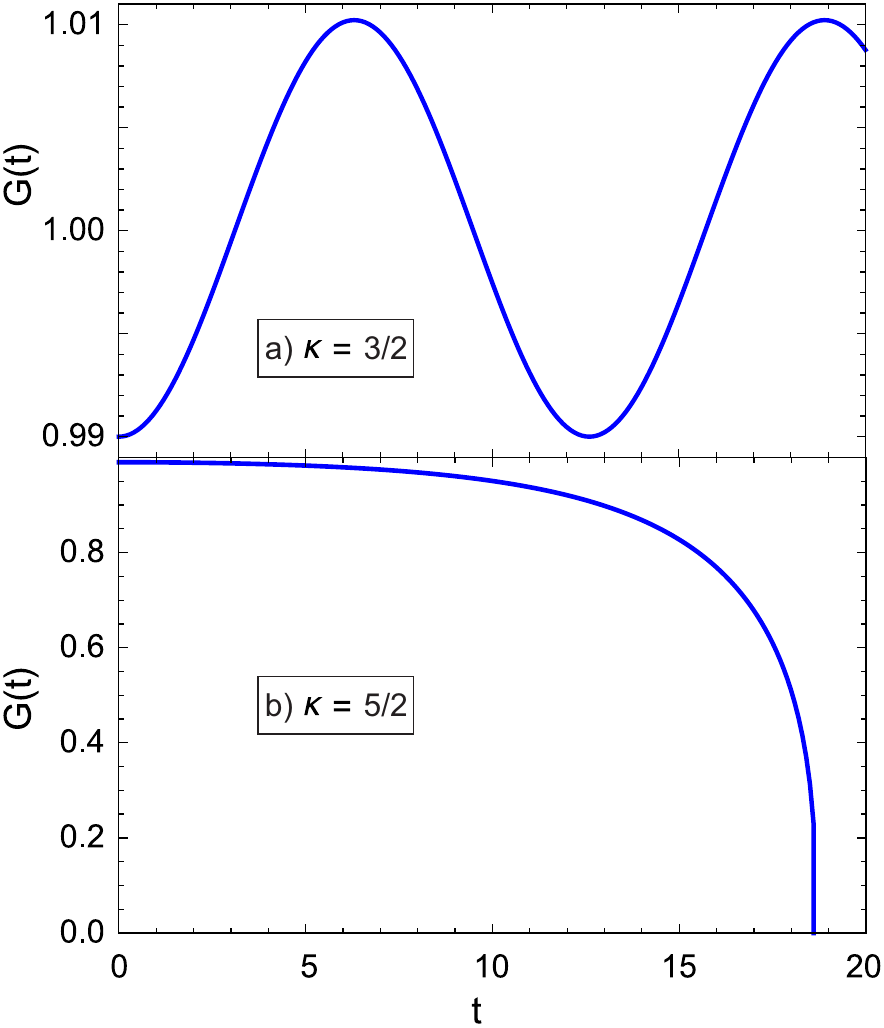}
  \caption{\label{f:NLSE-3-5} $G(t)$ from Eq.~\ef{var.e:9} 
  for (a)~$\kappa = 3/2$ and (b)~$\kappa = 5/2$. The latter case corresponds to ``blowup".}
\end{figure}

By multiplying both sides of \ef{var.e:9} by $\dot G$ and integrating with respect to time we obtain a first integral of the second order differential equation, which up to a multiplicative factor is the same as setting the conserved Hamiltonian divided by the mass $M$ to a constant $E$.  This gives
\begin{equation}\label{var.e:10}
   E 
   =    
   \frac{C_2}{C_1} \frac{ \dot G ^2}{4}
   +
   \frac{C_3}{C_1} \frac{1}{ G^2 }   
   - 
   \frac{1}{(\kappa+1)} \frac{2 \kappa C_3}{C_1} \,
   \qty( \frac{M}{C_1 G} )^\kappa \>.
\end{equation}
We notice that at the critical value of  $\kappa =2$, the last two terms both go like $1/G^2$. Self-focusing occurs when the width can go to zero.  Since $\dot G^2 $ needs to be positive, this means that at $\kappa=2$, the mass has to be greater than $M^\star$
for $G$ to be able to go to zero.  We find \cite{NLDE}
\begin{equation}\label{var.e:11}
    M^\star 
    = 
    \sqrt{ \frac{3 C_1^2} {4} }
    =  
    \frac{\pi}{2}\sqrt{3} =  2.7207 \dotsb \>,
\end{equation}
provided we use the exact solution (which is a zero-energy solution) for 
$\kappa=2$, namely  $f=\sech^{1/2}(z)$.  This agrees well with numerical estimates of the critical mass \cite{Rose:p} and is slightly lower than the variational estimate obtained earlier by Cooper \etal\ \cite{variational} using a post-Gaussian trial wave functions instead of a trial wave function based on the exact solution.  For $\kappa \neq 2$, if we use the mass of the exact solitary wave solution, the energy conservation equation \ef{var.e:10} simplifies to 
\begin{equation}
   E 
   =    
   \frac{C_2}{C_1} \, \frac{ \dot G ^2}{4}+\frac{C_3}{C_1}\,
   \qty( \frac{1}{ G^2 } - \frac{2}{\kappa G^\kappa} )
\end{equation} 
In the supercritical case when $G \to 0$, we have
\begin{equation}\label{var.e:12}
   \frac{C_2}{C_1} \, \frac{  \dot G ^2}{4}
   = 
   \frac{1}{(\kappa+1)} \frac{2 \kappa C_3}{C_1} \,
   \qty( \frac{M }{C_1 G} )^\kappa \>.   
\end{equation}
This ``mean-field'' result was obtained earlier in Refs.~\cite{var2,variational}. 
To show the difference between the stability at $\kappa=3/2$ and $\kappa=5/2$, we have solved Eq.~\ef{var.e:9} for the initial conditions $G(0)=0.001 $, $\dot{G}(0) = 0$, with the results shown in Fig.~\ref{f:NLSE-3-5}.

For small oscillations we can assume
\begin{equation}\label{var.e:13}
   G(t) = 1 + \epsilon \, g(t) \>,
\end{equation}
from which we obtain the equation,
\begin{gather}\label{var.e:14}
   \ddot{g} 
   + 
   \omega^2 \, g = 0 \>,
   \\
   \omega^2
   =
   \qty( C_3 / C_2 ) \,
   \qty[ 12 - 4 \kappa^2 \qty( M/C_1 )^\kappa ] \>.
   \notag
\end{gather}
Setting $\omega=0$ in Eq.~\eqref{var.e:14}, leads to the same criterion for the critical mass when $\kappa=2$. 
The same equation gives the frequency of small oscillations when $\kappa < 2$.  For $\kappa = 3/2$, the predicted period of oscillation is $T=2 \pi/\omega = 12.5998$ in good agreement with Fig.~\ref{f:NLSE-3-5}a. 
%
%
%
%
%
%

%
%
\begin{figure}[t]
   \centering
   \includegraphics[width=0.9\columnwidth]{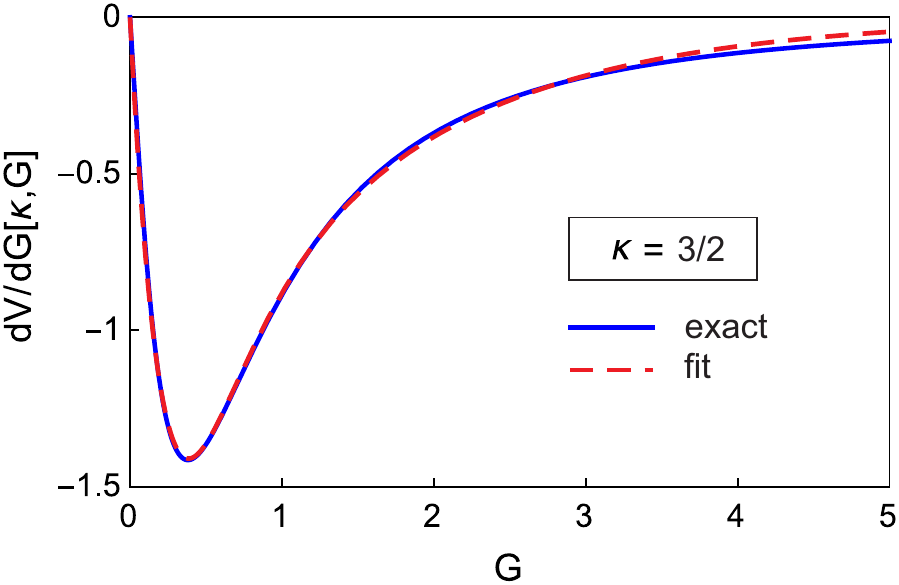}
   \caption{\label{f:Vfit} Potential fit for $\kappa = 3/2$. The solid line (blue online) 
   is the exact derivative of the potential from Eq.~\ef{var.e:19}, 
   the dashed line (red online) is the 
   fitted function $F(a,b,c,d)$ of Eq.~\ef{var.e:20}.}
\end{figure}
%
%

%
%
\subsection{\label{s:varNLSEexternal}Adding an external potential}

\begin{figure}[t]
  \centering
  \includegraphics[width=0.87\columnwidth]{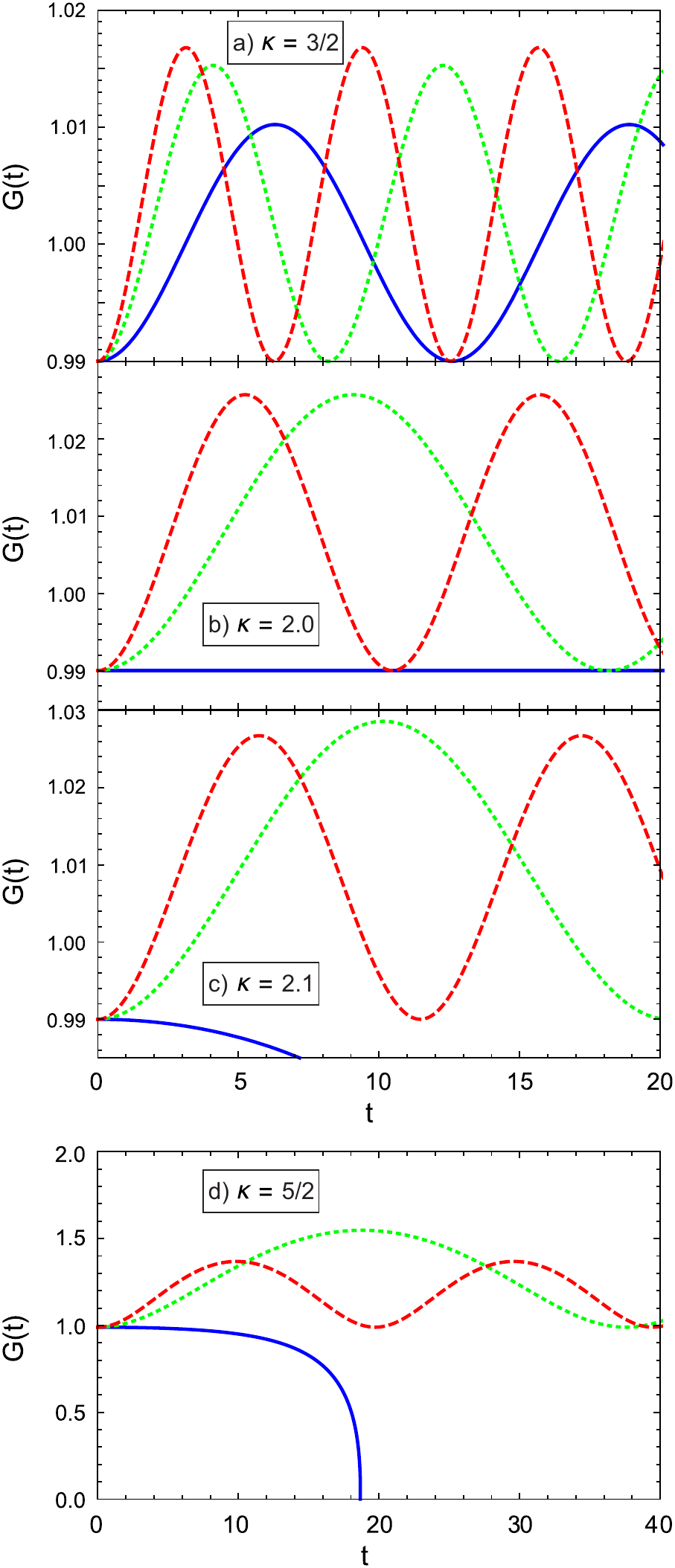}
  \caption{\label{f:Gt} The solid (blue online), dotted (green online), 
  and dashed (red online) lines are the solutions $G(t)$ of Eq.~\ef{var.e:9} 
  for $\kappa = 3/2$, $2$, $2.1$, and $5/2$, and 
  (a) $b^2 = 0.25,0.75,1.3611$, for $\kappa = 3/2$, 
  (b) $b^2 = 0.25,0.5,1$, for $\kappa = 2$, 
  (c) $b^2 = 0.25,0.5,0.95$, for $\kappa = 2.1$, and 
  (d) $b^2 = 0.25,0.5,0.81$, for $\kappa = 5/2$.}
\end{figure}

Now we would like to see how this argument is modified when we add the external potential $V^{-}(x)$.  In this case the exact solution is ``pinned'' to the origin.  The Lagrangian is again given by Eq.~\eqref{var.e:5} with the addition of the potential term:
\begin{align}\label{var.e:15}
   H'[G] / M
   &=
   \int_{-\infty}^{+\infty} \!\!\! \dd{x}
   \psi^{\ast}(x,t) \, V^{-}(x) \, \psi(x,t) / M
   \\
   &=
   - (b^2 - 1/4)\, \mc{V}[G,\kappa] / C_1 \>,
   \notag
\end{align}
where
\begin{align}\label{var.e:16}
   \mc{V}[G,\kappa]
   &=
   \int_{-\infty}^{+\infty} \!\!\! \dd{y}
   \sech^{2/\kappa}(y) \, \sech^{2}(G y) \>,
   \\
   &\xrightarrow{G \to 1}
   \frac{\sqrt{\pi} \, \Gamma[1+1/\kappa]}{\Gamma[3/2+1/\kappa]} \>.
   \notag
\end{align}
The Lagrangian now becomes:
\begin{align}
   &\frac{L[G,\Lambda]}{M}
   =
   \frac{1}{4} \, v^2
   +
   \omega
   -
   \dot{\Lambda} \, G^2 \, \frac{C_2}{C_1}
   -
   \frac{C_3}{C_1} \, \frac{1}{G^2}
   -
   4 \Lambda^2 \, \frac{C_2}{C_1} \, G^2
   \notag \\
   & \>
   +
   \frac{1}{(\kappa + 1)} \,  \frac{C_4}{C_1} \,
   \qty( \frac{M}{C_1 G} )^{\kappa}
   +
   \frac{(b^2 - 1/4)}{C_1} \, \mc{V}[G,\kappa] \>.
   \label{var.e:17}
\end{align}
The Euler-Lagrange equations now give
\begin{align}\label{var.e:18}
   \ddot{G} 
   &= 
   4 \, \frac{C_3}{C_2} \, \frac{1}{G^3}  
   - 
   \frac{4 \, \kappa^2}{(\kappa + 1)} \, 
   \frac{ C_3}{C_2 G} \, \qty( \frac{M}{C_1 G} )^{\kappa}
   \\
   & \hspace{2em}
   +
   \frac{2 \, (b^2 - 1/4)}{C_1} \,
   \pdv{\mc{V}[G,\kappa]}{G} \>,
   \notag
\end{align}
where
\begin{align}
   \pdv{\mc{V}[G,\kappa]}{G}
   &=
   - 2
   \int_{-\infty}^{+\infty} \!\!\! \dd{y}
   y \, \sech(Gy) \,  \sech^{3}(G y) \, \sech^{2/\kappa}(y)
   \notag \\
   &\xrightarrow{G \to 1}
   -
   \frac{\kappa}{\kappa + 1} \,
   \frac{\sqrt{\pi} \, \Gamma[1+1/\kappa]}{\Gamma[3/2+1/\kappa]} \>.
   \label{var.e:19}
\end{align}
To solve this equation numerically we fit the numerical values of the integral in \ef{var.e:19} by a function of the form:
\begin{equation}\label{var.e:20}
   F(a,b,c,d)
   =
   a \, e^{-d \, G} \sech^b(G) \tanh (c \, G) \>.
\end{equation}
Using Mathematica, one obtains an extremely accurate 4-parameter fit.  For example, the result of this fit for $\kappa = 3/2$ is shown in Fig.~\ref{f:Vfit} for $a=4.290$, $b=-1.528$, $c=-2.214$, and $d=2.222$.  Different fit parameters are used for each value of $\kappa$.  Plots of the solutions $G(t)$ of Eq.~\ef{var.e:18} for different values of $b$ and for $\kappa = 3/2$, $2$, $2.1$, and $5/2$ are shown in Fig.~\ref{f:Gt}.

%
%
%
%
%

We can study analytically the stability of the solutions in this variational approximation by linearizing Eq.~\ef{var.e:18} around the exact solution $G=1$,
\begin{equation}\label{var.e:21}
   G(t)
   = 1 + \epsilon \, g(t) \>.
\end{equation}
To evaluate the effect of the external potential on the small oscillation equation we just need to know that:
\begin{align}\label{var.e:22}
   &y \, \sech(Gy) \, \sech^{3}(G y)
   =
   y \, \sech(y) \, \sech^{3}(y) \\
   & \quad
   -
   \epsilon \, g(t) \, y^2
   \qty[
      2 \sech^2(y) - 3 \sech^4(y) ]
   +
   \order{ \epsilon^2 }
   \notag \>. 
\end{align}
Substitution of this expansion into \ef{var.e:18} gives
\begin{gather}\label{var.e:22a}
   \ddot{g} + \omega^2 \, g = 0 \>, \\
   \omega^2(\kappa,b)
   =
   (C_3/C_2) \,
   \qty[
      12
      -
      4 \kappa^2 \, \qty( b_{\text{max}}^2(\kappa) - b^2 ) ] 
      \notag \\
      \hspace{3em}
      -
      4 \, ( b^2 - 1/4 ) \, 
      \qty[ 2 \, ug(\kappa) - 3 \, ug_2(\kappa) ] / C_2 
      \notag \>, 
\end{gather}
with $b_{\text{max}}(\kappa) = 1/2 + 1/\kappa$, where $ug(\kappa)$ is given by Eq.~\ef{Dext.e:10} and where
\begin{equation}\label{var.e:23}
   ug_2(\kappa)
   =
   \int_{-\infty}^{+\infty} \!\!\! \dd{y}
   y^2 \, \sech^{2/\kappa + 4}(y) \>.
\end{equation}
%
%
%
%
The collective coordinate method allows one to approximately calculate the small oscillation frequency as well as the time evolution of the system using Eq.~\eqref{var.e:18}.
In Fig.~\ref{f:Gt} we assumed $\epsilon = 0.01, g(0)= -1 , \dot g = 0$.
The relevant values of $b_{\text{crit}}$ are $1/2, 0.525, 0.5785$ for $\kappa = 2, 2.1, 2.5$.  For $\kappa=3/2$ we get oscillation for the entire range from $b=1/2$ to $b=b_{max}$
as seen in Fig.~\ref{f:Gt}a.  As predicted, for $\kappa=2$, once we get above $b^2=1/4$, which is the case with no potential, then the solution is stable as seen in Fig.~\ref{f:Gt}b.  For $\kappa=2.1$, once we get above $b=b_{\text{crit}}$, then the solution is stable as seen in Fig.~\ref{f:Gt}c.  For $\kappa=5/2$ we get similar results to $\kappa= 2.1$, as seen in Fig.~\ref{f:Gt}d.  In the stable regime, the oscillation periods are accurately predicted by Eq.~\eqref{var.e:22a}. 

Setting $\omega^2(\kappa,b)=0$, determines the critical value of $b$ at a given $\kappa$ below which the solutions are unstable for $\kappa > 2$.  The expression for $b_{\text{crit}}$ obtained this way is identical to the expression for  $b_{\text{crit}}$ obtained from Derrick's theorem in Eq.~\eqref{Dext.e:13} and shown in Fig.~\ref{NS.f:bvskappa}.

\begin{figure}
\centering
\includegraphics[width=.9 \columnwidth]{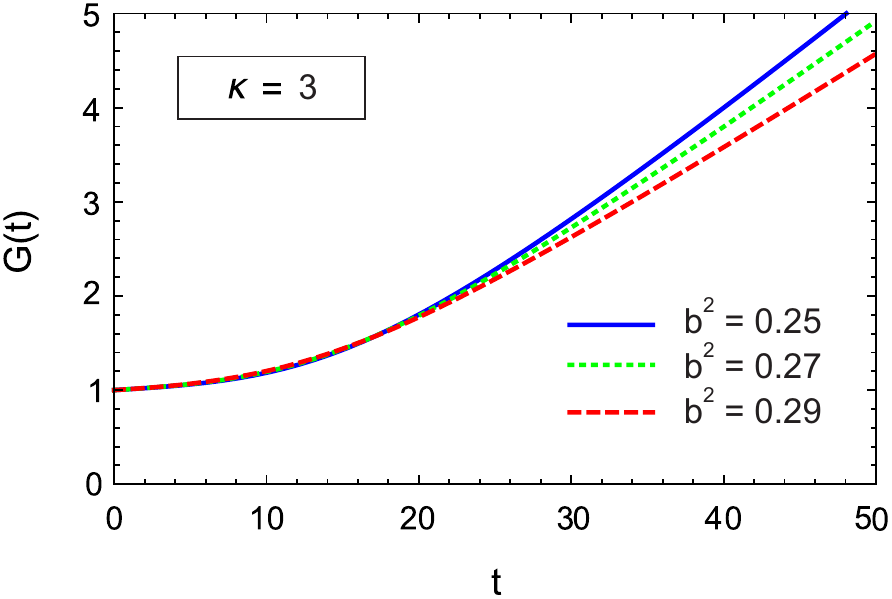}
\caption{$G(t)$   vs. $t$ at $\kappa=3$ for three values of $b^2=0.25, 0.27, 0.29.$  when $\dot G(0) =0.01$. This figure shows the collapse of the wave function for $\kappa=3$ in the unstable region,  $b < b_{crit}$.}
\label{collapse} 
\end{figure}

In the domain of instability one finds that if we look at initial conditions where $G(0)=1,  (g(0)=0)$ and $\epsilon=0.01$, $\dot g = \pm 1$, then for the minus sign one gets ``blow up'' ($G \rightarrow 0$), and for the plus sign we get collapse of the solution ($G \rightarrow \infty$).  In Fig.~\ref{collapse}, we give an example of collapse when $\kappa = 3$ and we are in the unstable regime.

%
%
%
%

A first integral of the second order differential equation resulting from the Lagrange's equation for $G$ can be obtained by setting the conserved Hamiltonian to a constant $E$.  One then has
\begin{align}
   E 
   &=   
   \frac{C_3}{C_1} \frac{1}{ G^2 }  
   +
   4 \Lambda^2 \, \frac{C_2}{C_1} \, G^2
   - 
   \frac{2 \kappa}{(\kappa+1)}\frac{C_3}{C_1} 
   \qty( \frac{M }{C_1 G})^\kappa 
   \notag \\
   & \hspace{2em}
   - 
   \frac{( b^2 - 1/4 )}{C_1 } \, \mc{V}[G,\kappa] \>.  
   \label{Econs}
\end{align}
From the energy conservation equation we can see immediately that at $\kappa=2$ the exact solution we found does not blow up.  This is for two reasons: first, when the width parameter $ G \rightarrow 0$, then $\mc{V}[G,\kappa]$ becomes a constant independent of $G$ and therefore
the potential does not affect the small $G$ behavior of the differential equation; secondly the mass of the exact solution depends now on $b$ and $\kappa$ and it is lower than the critical mass needed for blowup.  That is, the mass of the bound solution is given by:
\begin{align}
   M
   &=
   A^2[\kappa,b] \, C_1[\kappa]
   =
   \frac{\sqrt{\pi} \, 
         \qty( b_{\text{max}}^2(\kappa) - b^2 )^{1/\kappa} \,
         \Gamma[1/\kappa] }{ \Gamma[1/2 + 1/\kappa] }
    \notag \\
    & \hspace{2em}
    \xrightarrow{\kappa \to 2}
    \pi \, \sqrt{1 - b^2} \>.
    \label{Mkappav} 
\end{align}
The maximum value of this occurs when the external potential goes to zero at $b=1/2$.  When $b > 1/2$, the mass of the exact solution is \emph{always} less than $M^{\star}$, so that these solutions are always stable when $\kappa=2$.  For the NLSE with no external potential, when the stability depends on the mass of the initial wave function at $\kappa=2$, the critical value is that of the exact solitary wave solution. See also Fig.~\ref{NS.f:Mcrit} and the discussion thereof.

%
%
\section{\label{NS.s:LinearStability}Linear Stability} 

Let us perform the linear stability analysis of the solitary wave solutions $\phi_{\omega}(x) \, e^{-i\omega t}$ to the nonlinear \Schrodinger\ equation in the external potential.
We take a perturbed solitary wave solution in the form $\psi(x,t)= [\, \phi_{\omega}(x)+r(x,t) \, ] \, e^{-i\omega t}$ and consider the linearized equation on $R(x,t) = \qty[ \, \Re{r(x,t)},\Im{r(x,t)} \,]$,
\begin{equation}
   \partial_t R(x,t) = \mc{A}(\omega) \, R(x,t) \>.
\end{equation}
If the spectrum of $\mc{A}(\omega)$ has eigenvalues with positive real part, 
then the corresponding solitary wave is called linearly unstable; otherwise, 
it is called spectrally stable.  

In general, the spectral stability does not imply nonlinear stability, but for the nodeless solutions to the nonlinear \Schrodinger\ equation one 
can use the Lyapunov-type approach to prove the orbital stability; see 
e.g.~Ref.~\cite{MR901236}.

The equation we are solving is
\begin{equation}\label{nls-potential}
   \qty{
   i \,
   \partial_t
   +  
   \partial_x^2
   - 
   V^{-}(x)
   +
   | \psi(x,t) |^{2\kappa}
   } \, \psi(x,t) 
   = 
   0 \>,
\end{equation}
with
\begin{equation}\label{def-V}
   V^{-}(x)
   =
   -\qty(b^2-1/4) \, \sech^2{x} \>.
\end{equation}
We are interested in the stability of the solitary wave solution
$\psi_{\omega,b}(x,t) = \phi_{\omega,b}(x) \, e^{-i \omega t}$ to \eqref{nls-potential},
with the amplitude $\phi_{\omega,b}(x)$ satisfying
\begin{equation}\label{stationary-eq}
   \omega \, \phi_{\omega,b}(x)
   =
   \qty[ 
      -
      \partial_x^2
      +
      V\sp{-}(x)
      -
      \abs{\phi_{\omega,b}(x)}^{2\kappa}
       ] \, \phi_{\omega,b}(x)\>.
\end{equation}
For $\omega = \omega_{\kappa} = -1/\kappa^2$, one has the explicit expression
\begin{equation}\label{PF.e:psiphi-def}
    \phi_{\omega_{\kappa},b}(x)
    =
    \qty[ b_{\text{max}}^2(\kappa) - b^2 ]^{1/(2\kappa)} \,
    \sech^{1/\kappa}{x} \>, 
\end{equation}
with $b_{\text{max}}(\kappa) = 1/2 + 1/\kappa$.  We will perform the spectral analysis of the linearization operator following the V--K approach \cite{stab1}.  We consider the perturbation of the solitary wave, $\psi(x,t) = \qty[ \, \phi_{\omega,b}(x)+r(x,t) \, ] \, e^{-i\omega t}$,
with $r(x,t)=u(x,t)+iv(x,t)$, and with $u(x,t)$ and $v(x,t)$ real.
The linearized equation on $u(x,t)$ and $v(x,t)$ is given by
\begin{align}
   \partial_t \begin{pmatrix}u\\v\end{pmatrix}
   &=
   \mathcal{A}(\omega,b)\begin{pmatrix}u\\v\end{pmatrix}
   \label{def-operator-A} \\
   &=
   \begin{pmatrix}0&L_{-}(\omega,b)\\-L_{+}(\omega,b)&0\end{pmatrix} \,
   \begin{pmatrix}u\\v\end{pmatrix} \>,
   \notag
\end{align}
where the self-adjoint operators $L_\pm(\omega,b)$ are given by
\begin{subequations}\label{Lpmdefs}
\begin{align}
   L_{-}(\omega,b)
   &=
   - \partial^2_x
   +
   V^{-}(x) 
   -
   \omega
   -
   |\phi\sb{\omega,b}|^{2\kappa} \>,
   \label{Lpmdefs-a} \\
   L_{+}(\omega,b)
   &=
   - \partial^2_x
   +
   V^{-}(x) 
   -
   \omega
   -
   \qty( 2\kappa+1 ) \, |\phi\sb{\omega,b}|^{2\kappa} \>.
   \label{Lpmdefs-b}
\end{align}
\end{subequations}
The stationary equation \eqref{stationary-eq} satisfied by $\phi_{\omega,b}$ and its derivative with respect to $\omega$ give the relations
\begin{equation}\label{lm-lp}
   L_{-}(\omega,b)\,\phi_{\omega,b} = 0 
   \qc
   L_{+}(\omega,b) \, \partial_{\omega} \phi_{\omega,b} = \phi_{\omega,b} \>.
\end{equation}
We need to perform the spectral analysis of $L\sb{-}(\omega,b)$ and $L\sb{+}(\omega,b)$.

We start with reviewing the V--K approach from \cite{comech} for the case $b=1/2$, when $V^{-}(x)\equiv 0$.  For a given value $\kappa > 0$, let
\begin{align}\label{def-varphi}
   \varphi_\omega(x)
   &:=
   \phi\sb{\omega,1/2}(x)
   \\
   &=
   (\kappa+1)^{1/(2\kappa)} \,
   |\omega|^{1/(2\kappa)} \,
   \sech^{1/\kappa}(\kappa x\sqrt{|\omega|}) 
   \notag
\end{align}
be the profile of a solitary wave for the case when $V^{-}(x)=0$ (when $b=1/2$).
By the V--K theory, the linearization at $\varphi_\omega$ is such that
\begin{equation*}
   L_{-}(\omega) = L_{-}(\omega,1/2)
   =
   - \partial^2_x - \omega - |\varphi_\omega|^{2\kappa}
\end{equation*}
has a simple eigenvalue $\lambda = 0$ as its smallest eigenvalue, corresponding to the eigenfunction $\varphi_\omega(x)$, while
\begin{equation*}
   L_{+}(\omega)=   L_{+}(\omega,1/2)
   =
   - \partial^2_x - \omega - \qty( 2 \kappa + 1) \,|\varphi_\omega|^{2\kappa}
\end{equation*}
has one simple negative eigenvalue on the subspace of even functions, and a simple eigenvalue at $\lambda=0$ on the subspace of odd functions corresponding to the eigenfunction
$\partial_x \varphi_{\omega}(x)$.

For any nonzero eigenvalue $\lambda \in \sigma_{p}(\mc{A}(\omega,1/2))$
of the linearization operator from \eqref{def-operator-A}, one has the relation $\lambda^2\psi = -L_{-}(\omega)L_{+}(\omega) \, \psi$ with nonzero $\psi$.  Being in the range of $L_{-}$, which is self-adjoint, $\psi$ is orthogonal to the null space of $L_{-}(\omega)$; this allows us to arrive at
\begin{equation}\label{lambda-square}
   \lambda^2 \,
   \expval{\psi,L_{-}(\omega)^{-1}\psi}
   =
   -
   \expval{\psi,L_{+}(\omega)\psi} \>,
\end{equation}
hence $\lambda^2 \in \mathbb{R}$.  Thus, the linear instability could only be caused by a positive eigenvalue of $\mathcal{A}(\omega,1/2)$. From \eqref{lambda-square}, one can see that one could have $\lambda>0$ if the right-hand side of \eqref{lambda-square} becomes positive for some $\psi$ orthogonal to the kernel of $L_{-}(\omega)$; in other words, if the minimization problem
\begin{equation}\label{mp}
   \mu
   =
   \inf_{\substack{
            \expval{\psi,\varphi_{\omega} } = 0 , \\
            \expval{\psi,\psi} = 1 }}
   \expval{ \psi,L_{+}(\omega)\psi }
\end{equation}
gives a negative value of $\mu$.  By \cite{stab1}, finding the minimum of \eqref{mp}
under constraints $\expval{\psi,\psi} = 1$ and $\expval{\psi,\varphi_{\omega}} = 0$ leads to the relation
\begin{equation}\label{lm}
   L_{+}(\omega) \, \psi
   =
   \mu \, \psi + \nu \, \varphi\sb{\omega} \>,
\end{equation}
with $\mu,\,\nu$ Lagrange multipliers; pairing the above with $\psi$ shows that $\mu$ in \ef{mp} and \ef{lm} is the same.  Writing $\psi = (L_{+}(\omega)-\mu)^{-1} \, \nu \, \varphi_{\omega}$ and taking into account that $\expval{\psi,\varphi_{\omega}} = 0$, we see that we need to analyze the location of the first root of the V--K function
\begin{equation}\label{vk-function}
   f(z)
   =
   \expval{ \varphi_{\omega}, (L_{+}(\omega) - z)^{-1} \, \varphi_{\omega} } \>, 
\end{equation}
which is defined for $z$ in the resolvent set of the operator $L\sb{+}(\omega)$ restricted onto the subspace of even functions.  This domain includes the interval $(z_0,z_2)$, where $z_0 < 0$ is the smallest negative eigenvalue of $L_{+}$ and $z_2 > 0$ is the next eigenvalue of $L_{+}$,
on the subspace of even functions.  Since clearly $f'(z) > 0$ for $z \in (z_0,z_2)$, one has $f(\mu) = 0$ at some $\mu \in (z_0,z_2)$, $ \mu > 0$ (hence stability) if and only if $f(0) <0 $, which leads to $\expval{ \varphi\sb{\omega},L\sb{+}(\omega)^{-1}\varphi\sb{\omega} } < 0$,
and, using \eqref{lm-lp}, we arrive at the V--K stability condition
\begin{equation}\label{vk-condition}
   \dv{\omega}
   \expval{ \varphi_{\omega},\varphi_{\omega} } < 0 \>.
\end{equation}
An elementary computation based on \ef{def-varphi} shows that \ef{vk-condition} is satisfied
(for all $\omega<0$) if and only if $\kappa \in (0,2)$.  The left-hand side of \ef{vk-condition} becomes identically zero for $\kappa = 2$ and becomes positive for $\kappa > 2$ (again, for all $\omega<0$).

Now let us consider $L\sb\pm(\omega,b)$ with $b > 1/2$ and $\omega = \omega_{\kappa} = -1/\kappa^2$.  As in the case of no potential, one has $L_{-}(\omega,b)\ge 0$, with $\lambda = 0$ a simple eigenvalue corresponding to the eigenfunction $\phi\sb{\omega,b}$.
At $\omega = \omega_\kappa$, one has
\begin{equation*}
   |\phi_{\omega,b}(x)|^{2\kappa} - V^{-}(x)
   =
   |\varphi_\omega(x)|^{2\kappa}
\end{equation*}
and
\begin{equation*}
   L_{-}(\omega,b)
   =
   - \partial^2_x - \omega - |\varphi\sb{\omega}(x)|^{2\kappa}
   =
   L_{-}(\omega,1/2) \>.
\end{equation*}
We note that
\begin{equation*}
   L_{+}(\omega,b)
   =
   L_{-}(\omega,b) - 2\kappa \, |\phi_{\omega,b}(x)|^{2\kappa}
   <
   L_{-}(\omega,b) \>,
\end{equation*}
hence the smallest eigenvalue $z_0(\omega,b)$ of $L_{+}(\omega,b)$ (assumed on the subspace of even functions) is negative.  At $\omega = \omega_{\kappa} = - 1/\kappa^2$, one has
\begin{equation}\label{lp-kappa}
   L_{+}(\omega_{\kappa},b)
   =
   L_{+}(\omega_{\kappa},1/2) + 2\kappa \, (b^2-1/4) \, \sech^{2}{x} \>,
\end{equation}
hence for $b > b'$ and $b,\,b' \in (1/2,b_{\text{max}}(\kappa))$,
\begin{equation}\label{lp-greater}
   L_{+}(\omega_{\kappa},b) > L_{+}(\omega_{\kappa},b') \>. 
\end{equation}
Just as in the case $b=1/2$ which we considered above, the linear instability takes place
when the minimization problem
\begin{equation}\label{mp-b}
   \mu
   =
   \inf_{\substack{
            \expval{\psi,\varphi_{\omega} } = 0 , \\
            \expval{\psi,\psi} = 1 }}
   \expval{ \psi,L_{+}(\omega,b)\psi }
\end{equation}
gives a negative value of $\mu$.  As we already pointed out in the case $b=1/2$, one has $\mu > 0$ for $\kappa \in (0,2)$ (equivalently, $\varphi_{\omega}$ are linearly stable),
and $\mu=0$ for $\kappa=2$.  Due to \ef{lp-greater}, one then also has $\mu > 0$ for $\kappa\in (0,2)$, $\omega = -1/\kappa^2$, $b \in [\, 1/2,b_{\text{max}}(\kappa) \,)$ and for $\kappa=2$,
$\omega=-1/\kappa^2$, $b \in (\, 1/2,b_{\text{max}}(\kappa) \,)$.  Thus, for these values of $\kappa$ and $b$, the solitary waves $\phi\sb{\omega,b}e^{-\omega t}$ are spectrally stable.

For  $\kappa>2$, the story is different: while $\mu$ in \eqref{mp-b} is negative for $b=1/2$
corresponding to the linear instability of $\varphi\sb\omega(x) e^{-i\omega t}$, $\mu$ could become positive if $b$ exceeds some critical value $b_{\text{crit}}(\kappa)$:
\begin{equation*}
   \partial_{\omega} \!
   \expval{ \phi\sb{\omega,b},\phi\sb{\omega,b} } < 0
   \qc\!
   b \in (b_{\text{crit}}(\kappa),b_{\text{max}}(\kappa))
   \qc\!
   \omega=-1/\kappa^2.
\end{equation*}
Numerically, we proceed as follows. We pick $\kappa>2$ and use the shooting method to construct a solitary wave $\phi_{\omega,b}$ and find the critical value $b_{\text{crit}}>1/2$ above which $\partial_\omega \expval{\phi_{\omega,b},\phi_{\omega,b} }|_{(\omega_{\kappa},b)}$ becomes negative [that is, when $b = b_{\text{crit}}(\kappa)$, a positive eigenvalue from the spectrum of $\mathcal{A}(\omega,b)$ collides with a negative eigenvalue, and they produce a pair of purely imaginary eigenvalues; for $b \in (b_{\text{crit}}(\kappa), b_{\text{max}}(\kappa))$, spectral stability takes place].  This gives us the critical values $b_{\text{crit}}$ \vs\ $\kappa$ in agreement with Fig.~\ref{NS.f:bvskappa}.

We find remarkably that Derrick's theorem and the V--K spectral analysis of stability give identical results. The same result for the stability regime was also obtained by
setting the oscillation frequency for small oscillations around the exact solution to zero using the time dependent variational method.  In distinction with the case without a potential, in the presence of the  external potential $V^{-}(x)$ the results of the stability analysis are much more interesting because of the additional  $b$ dependence of the exact solution. For $\kappa >2$ it 
is possible to interpret the results of  V-K and Derrick's theorem in terms of a critical mass $M_{crit}$ below which the solution is stable, or equally in terms of a critical depth $b$ for the confining potential above which the solution is stable.

%
%
\section{\label{s:conclude}Conclusions}

In this paper we studied the stability of the exact solution of the NLSE in a real P{\"o}schl-Teller potential which is the SUSY partner of a \emph{complex} $\mathcal{PT}$ symmetric potential studied previously \cite{pt1}.  Unlike the previous problem which required detailed numerical analysis for every value of the nonlinearity parameter $\kappa$, the real external potential problem here results in a Hamiltonian dynamical system amenable to several variational approaches to the stability problem, such as Derrick's theorem \cite{ref:derrick}, V--K theory \cite{stab1}, and a time dependent variational approach.  Using these methods we were able to show that for $\kappa >2$ the pinned solution has a region of stability that was not available to the solitary wave solution of the NLSE without an external potential.  The latter solutions are known to blow up in a finite time interval when perturbed appropriately.  The analytic result for the re-entry regime of stability found using Derrick's theorem was corroborated by a numerical study of spectral stability based on the V--K theory.  
This result is different from the result found numerically for the stability of the solution for the complex SUSY partner external potential $V^{+}(x)$.
The analysis of the stability of the solutions for $V^{+}(x)$ in \cite{pt1} showed a very complicated pattern.  Even for $\kappa=1$ there is a regime of instability as a function of $b$ for the nodeless solution.  At $\kappa=3$ all the solutions found for $V^{+}(x)$, were  unstable due to oscillatory instabilities.  Only for $\kappa < 2/3$ were the solutions stable.  In contrast, for the $V^{-}(x)$ potential we are able to address the stability question for all $\kappa$ \emph {analytically} and show that the effect of the external potential is to introduce a new domain of stability for all $\kappa > 2$, when compared to the stability of the related solitary wave solutions in the absence of an external potential.
The stability properties of the solutions of the NLSE in the presence of the partner potentials $V^{\pm}(x)$ are quite different from one another due to the dissipative versus conservative nature of these potentials.

%
%
\acknowledgments
F.C.\ would like to thank the Santa Fe Institute  and the Center for Nonlinear Studies at Los Alamos National Laboratory for their hospitality. 
A.K.\ is grateful to Indian National Science Academy (INSA) for awarding him INSA Senior Scientist position at Savitribai Phule Pune
University, Pune, India. 
The research of A.C.\ was carried out at the Institute for Information Transmission Problems, Russian Academy of Sciences at the expense of the Russian Foundation for Sciences (Project 14-50-00150).
B.M.\ and J.F.D.\ would like to thank the Santa Fe Institute for their hospitality.  
B.M.\ acknowledges support from the National Science Foundation through its employee IR/D program.
The work of A.S.\ was supported by the U.S.\ Department of Energy. 

%
%

%
%
%
%

%
%
\end{document}